\documentclass[aps,groupedaddress,showpacs,floatfix]{revtex4}
\usepackage{amssymb}
\usepackage{amsmath}
\usepackage{graphicx}
\DeclareMathOperator{\sgn}{sgn}
\DeclareMathOperator{\Ai}{Ai}
\begin{document}

\title{Two-point free energy distribution function in (1+1) directed polymers}

\author{Victor Dotsenko}

\affiliation{LPTMC, Universit\'e Paris VI, 75252 Paris, France}

\affiliation{L.D.\ Landau Institute for Theoretical Physics,
 119334 Moscow, Russia}

\date{\today}

\begin{abstract}
In this brief technical communication it is demonstrated how
using Bethe ansatz technique the explicit expression for the two-point free energy 
distribution function in (1+1) directed polymers
can be derived   in rather simple way. Obtained result is equivalent to the one
derived earlier by Prolhac and Spohn \cite{Prolhac-Spohn}.

\end{abstract}


\maketitle

\section{Introduction}

We consider the model of directed polymers described in terms of an elastic string $\phi(\tau)$
directed along the $\tau$-axes within an interval $[0,t]$ and defined by the Hamiltonian
\begin{equation}
   \label{1}
   H[\phi(\tau), V] = \int_{0}^{t} d\tau
   \Bigl\{\frac{1}{2} \bigl[\partial_\tau \phi(\tau)\bigr]^2
   + V[\phi(\tau),\tau]\Bigr\};
\end{equation}
where the disorder potential $V[\phi,\tau]$
is Gaussian distributed with a zero mean $\overline{V(\phi,\tau)}=0$
and 
\begin{equation}
 \label{1a}
\overline{V(\phi,\tau)V(\phi',\tau')} =  u \delta(\tau-\tau') \delta(\phi-\phi')
\end{equation}
The parameter $u$ describes the strength of the disorder.
The partition function with the fixed boundary condition, $\phi(t)=x$ is:
\begin{equation}
\label{2}
   Z(x) = \int_{\phi(0)=0}^{\phi(t)=x}
              {\cal D} \phi(\tau)  \;  \mbox{\Large e}^{- \beta H[\phi]}
\; = \; \exp\{- \beta F(x)\}
\end{equation}
Correspondingly, $F(x)$ is the free energy of the polymer which at time $t$
arrives to the point $x$. In the limit of large $t$ random free energy scales as 
\begin{equation}
 \label{3}
\beta F \; = \; \beta f_{0} t \; + \; \lambda f
\end{equation}
where $f_{0}$ is the trivial self-averaging contribution 
(which can be easily eliminated by simple redefinition of the total free energy and the 
partition function) and   
\begin{equation}
 \label{4}
\lambda = \frac{1}{2} (\beta^{5}u^{2} t)^{1/3} \propto t^{1/3}
\end{equation}
In the limit $t \to \infty$ the random quantity $f \sim 1$ in eq.(\ref{3})
is described by the  universal Tracy-Widom distribution 
\cite{KPZ-TW1a,KPZ-TW2,BA-TW2,LeDoussal1}.
The aim of the present brief communication is the  study the two-point free energy 
probability distribution function:
\begin{equation}
\label{5a}
W(f_{1}, f_{2}; x_{1}, x_{2}) \; = \; \lim_{t\to\infty} \;
\mbox{Prob}\bigl[f(x_{1}) \; > \; f_{1}; \; f(x_{2}) \; > \; f_{2}\bigr] 
\end{equation}
Some time ago the result for this function has been derived in terms of the Bethe ansatz
replica technique under a particular decoupling assumption \cite{Prolhac-Spohn}. 
Here I'm going to recompute this function, again in terms of the same general scheme of the 
Bethe ansatz approach but using somewhat different computational tricks (which do not require
any supplementary assumptions). 
Since this function depends only on the distance between the two points,
$x \equiv |x_{2} - x_{1}|$,
to simplify formulas I'll  consider the particular case:
$x_{1} = -\frac{1}{2} x$ and  $x_{2} = +\frac{1}{2} x$. In other words, instead of (\ref{5a}),
I'll concentrate on the probability distribution function defined as follows:
\begin{equation}
\label{5}
W(f_{1}, f_{2}; x) \; = \; \lim_{t\to\infty} \;
\mbox{Prob}\bigl[f(-x/2) \; > \; f_{1}; \; f(x/2) \; > \; f_{2}\bigr] 
\end{equation}
Just recently it has been proven \cite{Imamura-Sasamoto-Spohn} that
the result of the present calculations (see eqs.(\ref{51})-(\ref{52}) below) 
is  equivalent to the ones obtained 
earlier \cite{Prolhac-Spohn}. 

\section{Two point distribution function}

In terms of the partition function, eq.(\ref{2}), above the probability distribution function,
eq.(\ref{5}) can be defined as follows:
\begin{equation}
\label{6}
W(f_{1}, f_{2}; x) = \lim_{\lambda\to\infty}
\sum_{L=0}^{\infty} \sum_{R=0}^{\infty}
\frac{(-1)^{L}}{L!} \frac{(-1)^{R}}{R!}
\exp\bigl(\lambda L f_{1} + \lambda R f_{2}\bigr) \;
\overline{\Bigl[Z(-x/2)\exp\{\beta f_{0} t\}\Bigr]^{L} \, 
\Bigl[Z(x/2)\exp\{\beta f_{0} t\}\Bigr]^{R}}
\end{equation}
Here, the averaging, denoted by $\overline{(...)}$ is performed over random
potentials (\ref{1a}). 
Performing the standard  averaging of the $(L+R)$-th power of the 
partition function, eq.(\ref{2}),  one gets
\begin{equation}
\label{7}
W(f_{1}, f_{2}; x) = \lim_{\lambda\to\infty}
\sum_{L=0}^{\infty} \sum_{R=0}^{\infty}
\frac{(-1)^{L}}{L!} \frac{(-1)^{R}}{R!}
\exp\bigl(\lambda L f_{1} + \lambda R f_{2} +\beta(L+R)f_{0}t\bigr) \;
\Psi\bigl(\underbrace{-x/2, ...,-x/2}_{L}, \underbrace{x/2, ..., x/2}_{R} ; \; t\bigr)
\end{equation}
where the time dependent wave function $\Psi(x_{1}, ..., x_{N}; t)$ is the solution of the 
imaginary time Schr\"odinger equation
\begin{equation}
 \label{8}
\beta \, \partial_t \Psi({\bf x}; t) \; = \; 
\Bigl[\frac{1}{2}\sum_{a=1}^{N}\partial_{x_a}^2
   +\frac{1}{2}\, \kappa \sum_{a\not=b}^{N} \delta(x_a-x_b)\Bigr]
\Psi({\bf x}; t)
\end{equation}
where $\kappa = \beta^{3}u$ and  the initial condition
\begin{equation}
   \label{9}
\Psi({\bf x}; t=0) = \Pi_{a=1}^{N} \delta(x_a)
\end{equation}

A generic  eigenstate of such system is characterized by $N$ momenta
$\{ q_{a} \} \; (a=1,...,N)$ which split into
$M$  ($1 \leq M \leq N$) "clusters" described by
continuous real momenta $q_{\alpha}$ $(\alpha = 1,...,M)$
and having $n_{\alpha}$ discrete imaginary "components"
\begin{equation}
   \label{10}
q_{a} \; \equiv \; q^{\alpha}_{r} \; = \;
q_{\alpha} - \frac{i\kappa}{2}  (n_{\alpha} + 1 - 2r)
\;\; ; \; \;\;\; \;\;\; \;\;\;
(r = 1, ..., n_{\alpha})
\end{equation}
with the global constraint
\begin{equation}
   \label{11}
\sum_{\alpha=1}^{M} n_{\alpha} = N
\end{equation}
A generic time dependent solution  $\Psi({\bf x},t)$
of the Schr\"odinger equation (\ref{8}) with the initial conditions, eq.(\ref{9}),
can be represented in the form of the linear combination of the eigenfunctions
$\Psi_{\bf q}^{(M)}({\bf x})$:
\begin{equation}
   \label{12}
\Psi(x_{1}, ..., x_{N}; t) =
\sum_{M=1}^{N} \frac{1}{M!} \Biggl[\int {\cal D}^{(M)} ({\bf q},{\bf n})\Biggr] \;
|C_{M}({\bf q},{\bf n})|^{2} \;
\Psi^{(M)}_{{\bf q}}({\bf x})
{\Psi^{(M)}_{{\bf q}}}^{*}({\bf 0}) \;
\exp\bigl\{-E_{M}({\bf q},{\bf n}) t \bigr\}
\end{equation}
where we have introduced the notation
\begin{equation}
   \label{13}
\int {\cal D}^{(M)} ({\bf q},{\bf n}) \equiv
\prod_{\alpha=1}^{M} \Biggl[\int_{-\infty}^{+\infty} \frac{dq_{\alpha}}{2\pi} \sum_{n_{\alpha}=1}^{\infty}\Biggr]
{\boldsymbol \delta}\Bigl(\sum_{\alpha=1}^{M} n_{\alpha} \; , \;  N\Bigr)
\end{equation}
and ${\boldsymbol \delta}(k , m)$ is the Kronecker symbol.
For a given set of integers $\{M; n_{1}, ...., n_{M}\}$
the eigenfunctions $\Psi_{\bf q}^{(M)}({\bf x})$ can be represented as follows:
\begin{equation}
\label{14}
\Psi^{(M)}_{{\bf q}}({\bf x}) =
\sum_{{\cal P}}  \;
\prod_{a<b}^{N}
\Biggl[
1 +i \kappa \frac{\sgn(x_{a}-x_{b})}{q_{{\cal P}_a} - q_{{\cal P}_b}}
\Biggr] \;
\exp\Bigl[i \sum_{a=1}^{N} q_{{\cal P}_{a}} x_{a} \Bigr]
\end{equation}
where the summation goes over $N!$ permutations ${\cal P}$ of $N$ momenta $q_{a}$,
eq.(\ref{10}),  over $N$ particles $x_{a}$;
the normalization factor
\begin{equation}
   \label{15}
|C_{M}({\bf q}, {\bf n})|^{2} = \frac{\kappa^{N}}{N! \prod_{\alpha=1}^{M}\bigl(\kappa n_{\alpha}\bigr)}
\prod_{\alpha<\beta}^{M}
\frac{\big|q_{\alpha}-q_{\beta} -\frac{i\kappa}{2}(n_{\alpha}-n_{\beta})\big|^{2}}{
      \big|q_{\alpha}-q_{\beta} -\frac{i\kappa}{2}(n_{\alpha}+n_{\beta})\big|^{2}}
\end{equation}
and the eigenvalues:
\begin{equation}
E_{M}({\bf q},{\bf n}) \; = \;
\sum_{\alpha=1}^{M} 
\Bigl[
\frac{1}{2\beta} n_{\alpha} q_{\alpha}^{2}
- \frac{\kappa^{2}}{24\beta} n_{\alpha}^{3}
\Bigr]
+ \frac{\kappa^{2}}{24\beta} N
\label{16}
\end{equation}
The last term in the above expression is  the 
self-averaging part of the free energy; choosing $f_{0} = \kappa^{2}/(24\beta^{2})$
this term drops out of the further calculations.
Note also that according to the definition, eq.(\ref{14}), 
$\Psi^{(M)}_{{\bf q}}({\bf 0}) = N! $

Substituting eqs.(\ref{12})-(\ref{16}) into eq.(\ref{7}),
we get:
\begin{eqnarray}
\label{17}
W(f_{1}, f_{2}; x) &=& 1 + \lim_{\lambda\to\infty}
\sum_{L+R\geq 1}^{\infty} \;\frac{(-1)^{L+R}}{L! \, R!} \;
\mbox{\LARGE e}^{\lambda L f_{1} + \lambda R f_{2}}
\times
\\
\nonumber
\\
\nonumber
&\times&
\sum_{M=1}^{L+R} \frac{1}{M!}
\prod_{\alpha=1}^{M}
\Biggl[
\sum_{n_{\alpha}=1}^{\infty}
\int_{-\infty}^{+\infty} \frac{dq_{\alpha}}{2\pi \kappa n_{\alpha}} \kappa^{n_{\alpha}}
\mbox{\LARGE e}^{-\frac{t}{2\beta} n_{\alpha} q_{\alpha}^{2}
+ \frac{\kappa^{2} t}{24 \beta} n_{\alpha}^{3} }
\Biggr]
\; {\boldsymbol \delta}\Bigl(\sum_{\alpha=1}^{M} n_{\alpha} \; , \;  L+R\Bigr)
\; |\tilde{C}_{M}({\bf q}, {\bf n})|^{2}
\times
\\
\nonumber
\\
\nonumber
&\times&
\sum_{{\cal P}^{(L,R)}}  \sum_{{\cal P}^{(L)}} \sum_{{\cal P}^{(R)}} \;
\prod_{a=1}^{L} \prod_{c=1}^{R}
\Biggl[
\frac{q_{{\cal P}_a^{(L)}} - q_{{\cal P}_c^{(R)}} - i \kappa}{q_{{\cal P}_a^{(L)}} - q_{{\cal P}_c^{(R)}}}
\Biggr] 
\exp\Bigl[-\frac{i}{2} x \sum_{a=1}^{L} q_{{\cal P}_{a}^{(L)}} 
          +\frac{i}{2} x \sum_{c=1}^{R} q_{{\cal P}_{c}^{(R)}} \Bigr]
\end{eqnarray}
where
\begin{equation}
   \label{18}
|\tilde{C}_{M}({\bf q}, {\bf n})|^{2} \; = \;
\prod_{\alpha<\beta}^{M}
\frac{\big|q_{\alpha}-q_{\beta} -\frac{i\kappa}{2}(n_{\alpha}-n_{\beta})\big|^{2}}{
      \big|q_{\alpha}-q_{\beta} -\frac{i\kappa}{2}(n_{\alpha}+n_{\beta})\big|^{2}}
\end{equation}
In eq.(\ref{17}) the summation over all permutations ${\cal P}$ of $(L+R)$ momenta
$\{q_{1}, ..., q_{L+R}\}$  over $L$ "left" particles
$\{x_{1}, ..., x_{L}\}$
and $R$ "right" particles $\{y_{R}, ..., y_{1}\}$
split   into three parts: the permutations ${\cal P}^{(L)}$
of $L$ momenta (taken at random out of the total list $\{q_{1}, ..., q_{L+R}\}$)
over $L$ "left" particles, the permutations ${\cal P}^{(R)}$
of the remaining $R$ momenta over $R$ "right" particles, and
finally the permutations ${\cal P}^{(L,R)}$ (or the exchange) of the
momenta between the group $"L"$ and the group $"R"$. It is evident that due to the
symmetry of the expression in eq.(\ref{17}) with respect to the permutations 
${\cal P}^{(L)}$ and ${\cal P}^{(R)}$ the summations over these permutations give
just the factor $L! \, R!$. 

Further simplification comes from the following general property of the
Bethe ansatz wave function, eq.(\ref{14}). It has such structure that
for ordered particles positions (e.g. $x_{1} < x_{2} < ... < x_{N}$)
in the summation over permutations the momenta $q_{a}$ belonging
to the same cluster also remain ordered. In other words,
if we consider the momenta, eq.(\ref{10}), of a cluster $\alpha$,
$\{q_{1}^{\alpha}, q_{2}^{\alpha}, ..., q_{n_{\alpha}}^{\alpha}\}$,
belonging to the particles $\{x_{i_{1}} < x_{i_{2}} < ... < x_{i_{n_{\alpha}}}\}$,
the permutation of any two momenta $q_{r}^{\alpha}$
and $q_{r'}^{\alpha}$ of this {\it ordered} set gives zero contribution.
Thus, in order to perform the summation over the permutations ${\cal P}^{(L,R)}$
in eq.(\ref{17}) it is sufficient to split the momenta of each cluster into two parts:
$\{q_{1}^{\alpha}, ...,  q_{m_{\alpha}}^{\alpha} ||
q_{m_{\alpha}+1}^{\alpha}..., q_{n_{\alpha}}^{\alpha}\}$, where $m_{\alpha} = 0, 1, ..., n_{\alpha}$ and
where the momenta $q_{1}^{\alpha}, ...,  q_{m_{\alpha}}^{\alpha}$ belong to the particles
of the sector $"L"$ (whose coordinates are all equal to $-x/2$), 
while the momenta $q_{m_{\alpha}+1}^{\alpha}..., q_{n_{\alpha}}^{\alpha}$
belong to the particles of the sector $"R"$ (whose coordinates are all equal to $+x/2$).

Let us introduce the numbering of the momenta
of the sector $"R"$ in the reversed order:
\begin{eqnarray}
\nonumber
q_{n_{\alpha}}^{\alpha} &\to&  {q^{*}}_{1}^{\alpha}
\\
\nonumber
q_{n_{\alpha}-1}^{\alpha} &\to&  {q^{*}}_{2}^{\alpha}
\\
\nonumber
&........&
\\
q_{m_{\alpha}+1}^{\alpha} &\to&  {q^{*}}_{s_{\alpha}}^{\alpha}
\label{19}
\end{eqnarray}
where $m_{\alpha} + s_{\alpha} = n_{\alpha}$ and (s.f. eq.(\ref{10}))

\begin{equation}
\label{20}
{q^{*}}_{r}^{\alpha} \; = \; q_{\alpha} + \frac{i \kappa}{2} (n_{\alpha} + 1 - 2r)
\; = \; q_{\alpha} + \frac{i \kappa}{2} (m_{\alpha} + s_{\alpha} + 1 - 2r)
\end{equation}
By definition, the integer parameters $\{m_{\alpha}\}$ and $\{s_{\alpha}\}$
fulfill the global constrains
\begin{eqnarray}
\label{21}
\sum_{\alpha=1}^{M} m_{\alpha} &=& L
\\
\nonumber
\\
\sum_{\alpha=1}^{M} s_{\alpha} &=& R
\label{22}
\end{eqnarray}
In this way the summation over permutations ${\cal P}^{(L,R)}$
in eq.(\ref{17}) is changed by the summations over the integer parameters
$\{m_{\alpha}\}$ and $\{s_{\alpha}\}$, 
which allows to lift the summations over $L$, $R$, and $\{n_{\alpha}\}$.
Straightforward  calculations result in the following
expression:
\begin{eqnarray}
 \nonumber
W(f_{1}, f_{2}; x) &=& \lim_{\lambda\to\infty}
\Biggl\{
1 + \sum_{M=1}^{\infty} \; \frac{(-1)^{M}}{M!} \;
\prod_{\alpha=1}^{M}
\Biggl[
\sum_{m_{\alpha}+s_{\alpha}\geq 1}^{\infty}
(-1)^{m_{\alpha}+s_{\alpha}-1}
\int_{-\infty}^{+\infty} \;
\frac{dq_{\alpha}}{
2\pi \kappa (m_{\alpha}+s_{\alpha})}
\times
\\
\nonumber
\\
\nonumber
&\times&
\exp\Bigl\{
\lambda m_{\alpha} f_{1} + \lambda s_{\alpha} f_{2}
-\frac{i}{2} x m_{\alpha} q_{\alpha} + \frac{i}{2} x s_{\alpha} q_{\alpha} -\frac{1}{2}\kappa x m_{\alpha} s_{\alpha}
-\frac{t}{2\beta} (m_{\alpha}+s_{\alpha}) q_{\alpha}^{2} +
\frac{\kappa^{2} t}{24 \beta} (m_{\alpha}+s_{\alpha})^{3} 
\Bigr\}
\Biggr]
\times
\\
\nonumber
\\
&\times&
|\tilde{C}_{M}({\bf q}, {\bf m + s})|^{2}  \;
{\bf G}_{M} \bigl({\bf q}, {\bf m}, {\bf s}\bigr)
\Biggr\}
\label{23}
\end{eqnarray}
where
\begin{eqnarray}
\nonumber
{\bf G}_{M} \bigl({\bf q}, {\bf m}, {\bf s}\bigr) 
&=&
\prod_{\alpha=1}^{M} \prod_{r=1}^{m_{\alpha}} \prod_{r'=1}^{s_{\alpha}}
\Biggl(
\frac{q^{\alpha}_{r} - {q^{*}}_{r'}^{\alpha} - i \kappa}{q^{\alpha}_{r} - {q^{*}}_{r'}^{\alpha}}
\Biggr)
\times
\prod_{\alpha<\beta}^{M} \prod_{r=1}^{m_{\alpha}} \prod_{r'=1}^{s_{\alpha}}
\Biggl(
\frac{q^{\alpha}_{r} - {q^{*}}_{r'}^{\beta} - i \kappa}{q^{\alpha}_{r} - {q^{*}}_{r'}^{\beta}}
\Biggr)
\\
\nonumber
\\
\label{24}
&=& 
\prod_{\alpha=1}^{M} 
\frac{
\Gamma\bigl(1 + m_{\alpha} + s_{\alpha}\bigr)}{
\Gamma\bigl(1 + m_{\alpha}\bigr) \Gamma\bigl(1 + s_{\alpha}\bigr)}
\times
\\
\nonumber
\\
\nonumber
&\times&
\prod_{\alpha<\beta}^{M} 
\frac{
\Gamma
\Bigl[
1 + \frac{m_{\alpha} + m_{\beta} + s_{\alpha} + s_{\beta}}{2}
+\frac{i}{\kappa}\bigl(q_{\alpha} - q_{\beta}\bigr)
\Bigr] \,
\Gamma
\Bigl[
1 + \frac{-m_{\alpha} + m_{\beta} + s_{\alpha} - s_{\beta}}{2}
+\frac{i}{\kappa}\bigl(q_{\alpha} - q_{\beta}\bigr)
\Bigr]}{
\Gamma
\Bigl[
1 + \frac{-m_{\alpha} + m_{\beta} + s_{\alpha} + s_{\beta}}{2}
+\frac{i}{\kappa}\bigl(q_{\alpha} - q_{\beta}\bigr)
\Bigr] \,
\Gamma
\Bigl[
1 + \frac{m_{\alpha} + m_{\beta} + s_{\alpha} - s_{\beta}}{2}
+\frac{i}{\kappa}\bigl(q_{\alpha} - q_{\beta}\bigr)
\Bigr]}
\end{eqnarray}
After rescaling
\begin{eqnarray}
\label{25}
q_{\alpha} &\to& \frac{\kappa}{2\lambda} \, q_{\alpha}
\\
\nonumber
\\
\label{26}
x &\to&  \frac{2 \lambda^{2}}{\kappa} \, x
\end{eqnarray}
with
\begin{equation}
\label{27}
\lambda \; = \; \frac{1}{2} \,
\Bigl(\frac{\kappa^{2} t}{\beta}\Bigr)^{1/3} \; = \;
\frac{1}{2} \, \bigl(\beta^{5} u^{2} t\bigr)^{1/3}
\end{equation}
the normalization factor $|\tilde{C}_{M}({\bf q}, {\bf m + s})|^{2}$, eq.(\ref{18}),
can be represented as follows:
\begin{eqnarray}
\nonumber
|\tilde{C}_{M}({\bf q}, {\bf m + s})|^{2} &=&
\prod_{\alpha<\beta}^{M}
\frac{
\big|
\lambda\bigl(m_{\alpha}+s_{\alpha}\bigr) - \lambda\bigl(m_{\beta}+s_{\beta}\bigr) -
i p_{\alpha} + ip_{\beta}
\big|^{2}}{
\big|
\lambda\bigl(m_{\alpha}+s_{\alpha}\bigr) + \lambda\bigl(m_{\beta}+s_{\beta}\bigr) -
i p_{\alpha} + ip_{\beta}
\big|^{2} }
\; = \;
\\
\nonumber
\\
\label{28}
&=&
\Biggl(\prod_{\alpha=1}^{M}
\bigl[2\lambda \bigl(m_{\alpha}+s_{\alpha}\bigr)\bigr]\Biggr)
\times
\det
\Biggl[
\frac{1}{
\lambda\bigl(m_{\alpha}+s_{\alpha}\bigr) - ip_{\alpha} +
\lambda\bigl(m_{\beta}+s_{\beta}\bigr) + ip_{\beta}}
\Biggr]_{\alpha,\beta=1,...,M}
\end{eqnarray}
Substituting eqs.(\ref{25})-(\ref{28}) into eq.(\ref{23}) and using the Airy function relation
\begin{equation}
   \label{29}
\exp\Bigl[ \frac{1}{3} \lambda^{3} (m_{\alpha}+s_{\alpha})^{3} \Bigr] \; = \;
\int_{-\infty}^{+\infty} dy \; \Ai(y) \;
\exp\Bigl[\lambda (m_{\alpha}+s_{\alpha}) \, y \Bigr]
\end{equation}
we get
\begin{eqnarray}
 \nonumber
W(f_{1},f_{2};x) \; = \; \lim_{\lambda\to\infty}
\Biggl\{
&1& + \sum_{M=1}^{\infty} \; \frac{(-1)^{M}}{M!} \;
\prod_{\alpha=1}^{M}
\Biggl[
\int\int_{-\infty}^{+\infty} \frac{dy_{\alpha} dq_{\alpha}}{2\pi}
\Ai\bigl(y_{\alpha} + q_{\alpha}^{2}\bigr)
\sum_{m_{\alpha}+s_{\alpha}\geq 1}^{\infty} (-1)^{m_{\alpha}+s_{\alpha}-1}
\times
\\
\nonumber
\\
\nonumber
&\times&
\exp\Bigl\{
\lambda m_{\alpha} (y_{\alpha} + f_{1} - \frac{1}{2} i q_{\alpha} x) +
\lambda s_{\alpha} (y_{\alpha} + f_{2} + \frac{1}{2} i q_{\alpha} x)
- \lambda^{2} m_{\alpha} s_{\alpha} x
\Bigr\} \;
\Biggr]
\times
\\
\nonumber
\\
&\times&
\det \hat{K}\Bigl[(\lambda m_{\alpha},\, \lambda s_{\alpha}, \, q_{\alpha});
(\lambda m_{\beta}, \, \lambda s_{\beta}, \, q_{\beta})\Bigr]_{\alpha,\beta=1,...,M}
\;
{\bf G}_{M} \Bigl(\frac{{\kappa \bf q}}{2\lambda}, \; {\bf m}, \; {\bf s}\Bigr)
\Biggr\}
\label{30}
\end{eqnarray}
where
\begin{equation}
\label{31}
\hat{K}\Bigl[(\lambda m, \, \lambda s, \, q); (\lambda m', \, \lambda s', \, q')\Bigr]
\; = \;
\frac{1}{
\lambda m + \lambda s - iq +
\lambda m' + \lambda s' + iq'}
\end{equation}
Using the relation
\begin{equation}
 \label{32}
\exp\{- \lambda^{2} m s x \} \; = \; 
\int_{-\infty}^{+\infty}
\frac{d\xi_{1}d\xi_{2}d\xi_{3}}{(2\pi)^{3/2}} 
\exp\Bigl\{
-\frac{1}{2} \xi_{1}^{2} -\frac{1}{2} \xi_{2}^{2}-\frac{1}{2} \xi_{3}^{2}
+\lambda m \sqrt{x}\xi_{1} + \lambda s \sqrt{x} \xi_{2} + i\lambda (m+s) \sqrt{x} \xi_{3}
\Bigr\}
\end{equation}
the expression in eq.(\ref{30}) can be represented as follows:
\begin{eqnarray}
 \nonumber
W(f_{1},f_{2};x) \; = \; 
&1& + \sum_{M=1}^{\infty} \; \frac{(-1)^{M}}{M!} \;
\prod_{\alpha=1}^{M}
\Biggl[
\int_{-\infty}^{+\infty} \frac{dy_{\alpha} dq_{\alpha}}{2\pi}
\frac{d{\xi_{1}}_{\alpha} d{\xi_{2}}_{\alpha}d{\xi_{3}}_{\alpha}}{(2\pi)^{3/2}} \; 
\Ai\bigl(y_{\alpha} + q_{\alpha}^{2} - i{\xi_{3}}_{\alpha} \sqrt{x}\bigr)
\times
\\
\nonumber
\\
\label{33}
&\times&
\exp\Bigl\{
-\frac{1}{2} \bigl({\xi_{1}}_{\alpha} + \frac{1}{2}iq_{\alpha}\sqrt{x}\bigr)^{2} 
-\frac{1}{2} \bigl({\xi_{2}}_{\alpha} - \frac{1}{2}iq_{\alpha}\sqrt{x}\bigr)^{2}
-\frac{1}{2} {\xi_{3}}_{\alpha}^{2}
\Bigr\}
\Biggr] \; 
{\cal S}_{M}\bigl({\bf q}, {\bf y}, {\boldsymbol \xi}_{1}, {\boldsymbol \xi}_{1},
 f_{1}, f_{2}, x\bigr)
\end{eqnarray}
where
\begin{eqnarray}
\nonumber
{\cal S}_{M}\bigl({\bf q}, {\bf y}, {\boldsymbol \xi}_{1}, {\boldsymbol \xi}_{1},
 f_{1}, f_{2}, x\bigr) 
&=&
\lim_{\lambda\to\infty}
\prod_{\alpha=1}^{M}
\Biggl[
\sum_{m_{\alpha}+s_{\alpha}\geq 1}^{\infty} (-1)^{m_{\alpha}+s_{\alpha}-1}
\exp\Bigl\{
\lambda m_{\alpha} (y_{\alpha} + f_{1} + {\xi_{1}}_{\alpha}\sqrt{x}) +
\lambda s_{\alpha} (y_{\alpha} + f_{2} + {\xi_{2}}_{\alpha}\sqrt{x})
\Bigr\} \;
\Biggr]
\times
\\
\nonumber
\\
&\times&
\det \hat{K}\Bigl[(\lambda m_{\alpha},\, \lambda s_{\alpha}, \, q_{\alpha});
(\lambda m_{\beta}, \, \lambda s_{\beta}, \, q_{\beta})\Bigr]_{\alpha,\beta=1,...,M}
\;
{\bf G}_{M} \Bigl(\frac{{\kappa \bf q}}{2\lambda}, \; {\bf m}, \; {\bf s}\Bigr)
\label{34}
\end{eqnarray}
To demonstrate how  the summations over $\{m_{\alpha}\}$
and $\{s_{\alpha}\}$ are performed in the limit $\lambda \to \infty$
let us consider the example of a general type:
\begin{equation}
\label{35}
R \; = \; \lim_{\lambda\to\infty} \prod_{\alpha=1}^{M}
\Biggl[
\sum_{n_{\alpha}=1}^{\infty} \; (-1)^{n_{\alpha} - 1}
\exp\{\lambda n_{\alpha} y_{\alpha}\}
\Biggr]\;
\Phi\bigl(\lambda; \; n_{1}, \; ..., \; n_{M}\bigr)
\end{equation}
where $\Phi$ is a function which depends both on $\lambda$ and on all summation parameters
$\{n_{1}, \; ..., \; n_{M}\}$. The above summations can be represented in terms of the 
integrals in the complex plane: 
\begin{equation}
\label{36}
R \; = \; \lim_{\lambda\to\infty} \prod_{\alpha=1}^{M}
\Biggl[
\frac{1}{2i} \int_{{\cal C}} \frac{dz_{\alpha}}{\sin(\pi z_{\alpha})}
\exp\{\lambda z_{\alpha} y_{\alpha}\}
\Biggr]\;
\Phi\bigl(\lambda; \; z_{1}, \; ..., \; z_{M}\bigr)
\end{equation}
where the integration goes over the contour ${\cal C}$ shown in Fig.1(a).
Shifting the contour to the position ${\cal C}'$ shown in Fig.1(b)
(assuming that there is no contribution from infinity), 
and redefining $z_{\alpha} \to z_{\alpha}/\lambda$, in the
limit $\lambda \to \infty$ we get:
\begin{equation}
\label{37}
R \; = \;  \prod_{\alpha=1}^{M}
\Biggl[
\frac{1}{2\pi i} \int_{{\cal C}'} \frac{dz_{\alpha}}{z_{\alpha}}
\exp\{z_{\alpha} y_{\alpha}\}
\Biggr]\;
\lim_{\lambda\to\infty}
\Phi\bigl(\lambda; \; z_{1}/\lambda, \; ..., \; z_{M}/\lambda\bigr)
\end{equation}
where the parameters $y_{\alpha}$ and $z_{\alpha}$
remain finite in the limit $\lambda \to \infty$.

\begin{figure}[h]
\begin{center}
   \includegraphics[width=12.0cm]{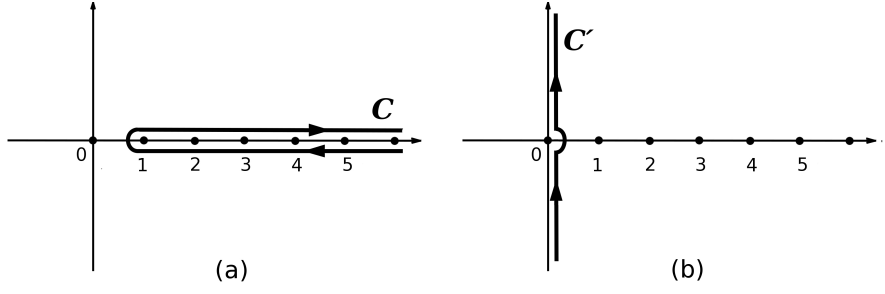}
\caption[]{The contours of integration in the complex plane used for
           summing the series:
           (a) the original contour ${\cal C}$;
           (b) the deformed contour ${\cal C}'$; }
\end{center}
\label{figure1}
\end{figure}
The double summations over $m_{\alpha}$ and $s_{\alpha}$ in eq.(\ref{34}) can
be represented as follows
\begin{equation}
\label{38}
\sum_{m_{\alpha}+s_{\alpha}\geq 1}^{\infty} (-1)^{m_{\alpha}+s_{\alpha}-1}
 =
\sum_{s_{\alpha}=0}^{\infty} \delta_{s_{\alpha}, 0}
\sum_{m_{\alpha}=1}^{\infty} (-1)^{m_{\alpha}-1}  +
\sum_{m_{\alpha}=0}^{\infty} \delta_{m_{\alpha}, 0}
\sum_{s_{\alpha}=1}^{\infty} (-1)^{s_{\alpha}-1}  -
\sum_{m_{\alpha}=1}^{\infty} (-1)^{m_{\alpha}-1}
\sum_{s_{\alpha}=1}^{\infty} (-1)^{s_{\alpha}-1}
\end{equation}
Thus in the integral representation, eqs.(\ref{35})-(\ref{37}),  for the function
in eq.(\ref{34}), we get
\begin{eqnarray}
\nonumber
{\cal S}_{M}\bigl({\bf q}, {\bf y}, {\boldsymbol \xi}_{1}, {\boldsymbol \xi}_{1},
 f_{1}, f_{2}, x\bigr) 
 &=&
\prod_{\alpha=1}^{M}
\Biggl[
\int\int_{{\cal C}'}
\frac{d{z_{1}}_{\alpha}d{z_{2}}_{\alpha}}{(2\pi i)^{2}}
\Bigl(
\frac{2\pi i}{{z_{1}}_{\alpha}} \delta({z_{2}}_{\alpha}) +
\frac{2\pi i}{{z_{2}}_{\alpha}} \delta({z_{1}}_{\alpha}) -
\frac{1}{{z_{1}}_{\alpha}{z_{2}}_{\alpha}}
\Bigr)
\times
\\
\nonumber
\\
&\times&
\exp\Bigl\{
{z_{1}}_{\alpha}\bigl(y_{\alpha} + f_{1} + {\xi_{1}}_{\alpha}\sqrt{x}\bigr) +
{z_{2}}_{\alpha}\bigl(y_{\alpha} + f_{2} + {\xi_{2}}_{\alpha}\sqrt{x}\bigr)
\Bigr\}
\Biggr] 
\times 
\\
\nonumber
\\
&\times&
\det \hat{K}\bigl[({z_{1}}_{\alpha},{z_{2}}_{\alpha},q_{\alpha});
({z_{1}}_{\beta},{z_{2}}_{\beta}, q_{\beta})\bigr]
\times
\Biggl[
\lim_{\lambda\to\infty}
{\bf G}_{M}\Bigl(
\frac{\kappa {\bf q}}{2\lambda}, \; 
\frac{{\bf z_{1}}}{\lambda}, \; 
\frac{{\bf z_{2}}}{\lambda}\Bigr)
\Biggr]
\label{39}
\end{eqnarray}
Using the explicit form of the factor ${\bf G}_{M}$, eq.(\ref{24}), and taking into account
the gamma function property $\lim_{|z|\to0} \Gamma(1+z) = 1$, we find
\begin{equation}
\label{40}
\lim_{\lambda\to\infty}
{\bf G}_{M}
\Bigl(\frac{\kappa {\bf q}}{2\lambda}, \; 
\frac{{\bf z_{1}}}{\lambda}, \; 
\frac{{\bf z_{2}}}{\lambda}\Bigr)
\; = \;
1
\end{equation}
Thus, in the limit $\lambda \to \infty$ the expression for the probability distribution function, eq.(\ref{33}),
takes the form of the Fredholm determinant
\begin{eqnarray}
 \nonumber
W(f_{1},f_{2};x) \; = \; 
&1& + \sum_{M=1}^{\infty} \; \frac{(-1)^{M}}{M!} \;
\prod_{\alpha=1}^{M}
\Biggl[
\int_{-\infty}^{+\infty} \frac{dy_{\alpha} dq_{\alpha}}{2\pi}
\frac{d{\xi_{1}}_{\alpha} d{\xi_{2}}_{\alpha}d{\xi_{3}}_{\alpha}}{(2\pi)^{3/2}} \; 
\Ai\bigl(y_{\alpha} + q_{\alpha}^{2} - i{\xi_{3}}_{\alpha} \sqrt{x}\bigr)
\times
\\
\nonumber
\\
\nonumber
&\times&
\exp\Bigl\{
-\frac{1}{2} \bigl({\xi_{1}}_{\alpha} + \frac{1}{2}iq_{\alpha}\sqrt{x}\bigr)^{2} 
-\frac{1}{2} \bigl({\xi_{2}}_{\alpha} - \frac{1}{2}iq_{\alpha}\sqrt{x}\bigr)^{2}
-\frac{1}{2} {\xi_{3}}_{\alpha}^{2}
\Bigr\} 
\times 
\\
\nonumber
\\
\nonumber
&\times&
\int\int_{{\cal C}'}
\frac{d{z_{1}}_{\alpha}d{z_{2}}_{\alpha}}{(2\pi i)^{2}}
\Bigl(
\frac{2\pi i}{{z_{1}}_{\alpha}} \delta({z_{2}}_{\alpha}) +
\frac{2\pi i}{{z_{2}}_{\alpha}} \delta({z_{1}}_{\alpha}) -
\frac{1}{{z_{1}}_{\alpha}{z_{2}}_{\alpha}}
\Bigr)
\times
\\
\nonumber
\\
\nonumber
&\times&
\exp\Bigl\{
{z_{1}}_{\alpha}\bigl(y_{\alpha} + f_{1} + {\xi_{1}}_{\alpha}\sqrt{x}\bigr) +
{z_{2}}_{\alpha}\bigl(y_{\alpha} + f_{2} + {\xi_{2}}_{\alpha}\sqrt{x}\bigr)
\Bigr\}
\Biggr] \; 
\times 
\\
\nonumber
\\
\label{41}
&\times&
\det \hat{K}\bigl[({z_{1}}_{\alpha},{z_{2}}_{\alpha},q_{\alpha});
({z_{1}}_{\beta},{z_{2}}_{\beta}, q_{\beta})\bigr]_{\alpha,\beta=1,...,M}
\\
\nonumber
\\
\nonumber
&=&
\det\bigl[\hat{1} \, - \, \hat{A} \bigr]
\end{eqnarray}
with the kernel
\begin{eqnarray}
 \nonumber
\hat{A}\bigl[({z_{1}}, \, {z_{2}}, \, q); ({z_{1}}', \, {z_{2}}', \, q')\bigr]
&=&
\int_{-\infty}^{+\infty} \frac{dy}{2\pi}
\frac{d{\xi_{1}} d{\xi_{2}}d{\xi_{3}}}{(2\pi)^{3/2}} \; 
\Ai\bigl(y + q^{2} - i{\xi_{3}} \sqrt{x}\bigr)
\times
\\
\nonumber
\\
\nonumber
&\times&
\exp\Bigl\{
-\frac{1}{2} \bigl({\xi_{1}} + \frac{1}{2}iq\sqrt{x}\bigr)^{2} 
-\frac{1}{2} \bigl({\xi_{2}} - \frac{1}{2}iq\sqrt{x}\bigr)^{2}
-\frac{1}{2} {\xi_{3}}^{2}
\Bigr\} 
\times 
\\
\nonumber
\\
\nonumber
&\times&
\Bigl(
\frac{2\pi i}{{z_{1}}} \delta({z_{2}}) +
\frac{2\pi i}{{z_{2}}} \delta({z_{1}}) -
\frac{1}{{z_{1}}{z_{2}}}
\Bigr)
\times
\\
\nonumber
\\
\nonumber
&\times&
\exp\Bigl\{
{z_{1}}\bigl(y + f_{1} + {\xi_{1}}\sqrt{x}\bigr) +
{z_{2}}\bigl(y + f_{2} + {\xi_{2}}\sqrt{x}\bigr)
\Bigr\} 
\times 
\\
\nonumber
\\
\label{42}
&\times&
\frac{1}{z_{1}+z_{2}-iq + z_{1}+z_{2}+iq}
\end{eqnarray}
In the exponential representation of this determinant we get
\begin{equation}
 \label{43}
W(f_{1},f_{2},x) \; = \;
\exp\Bigl[-\sum_{M=1}^{\infty} \frac{1}{M} \; \mbox{Tr} \, \hat{A}^{M} \Bigr]
\end{equation}
where
\begin{eqnarray}
 \nonumber
\mbox{Tr} \, \hat{A}^{M} &=&
\prod_{\alpha=1}^{M}
\Biggl[
\int_{-\infty}^{+\infty} \frac{dy dq_{\alpha}}{2\pi}
\frac{d{\xi_{1}} d{\xi_{2}}d{\xi_{3}}}{(2\pi)^{3/2}} \; 
\Ai\bigl(y + q_{\alpha}^{2} - i{\xi_{3}} \sqrt{x}\bigr)
\times
\\
\nonumber
\\
\nonumber
&\times&
\exp\Bigl\{
-\frac{1}{2} \bigl({\xi_{1}} + \frac{1}{2}iq_{\alpha}\sqrt{x}\bigr)^{2} 
-\frac{1}{2} \bigl({\xi_{2}} - \frac{1}{2}iq_{\alpha}\sqrt{x}\bigr)^{2}
-\frac{1}{2} {\xi_{3}}^{2}
\Bigr\} 
\times 
\\
\nonumber
\\
\nonumber
&\times&
\int\int_{{\cal C}'}
\frac{d{z_{1}}_{\alpha}d{z_{2}}_{\alpha}}{(2\pi i)^{2}}
\Bigl(
\frac{2\pi i}{{z_{1}}_{\alpha}} \delta({z_{2}}_{\alpha}) +
\frac{2\pi i}{{z_{2}}_{\alpha}} \delta({z_{1}}_{\alpha}) -
\frac{1}{{z_{1}}_{\alpha}{z_{2}}_{\alpha}}
\Bigr)
\times
\\
\nonumber
\\
\nonumber
&\times&
\exp\Bigl\{
{z_{1}}_{\alpha}\bigl(y + f_{1} + {\xi_{1}}\sqrt{x}\bigr) +
{z_{2}}_{\alpha}\bigl(y + f_{2} + {\xi_{2}}\sqrt{x}\bigr)
\Bigr\}
\Biggr]
\times
\\
\nonumber
\\
\label{44}
&\times&
\prod_{\alpha=1}^{M}
\Biggl[
\frac{1}{{z_{1}}_{\alpha} + {z_{2}}_{\alpha} - iq_{\alpha} 
       + {z_{1}}_{\alpha+1} + {z_{2}}_{\alpha+1} + iq_{\alpha+1}}
\Biggr]
\end{eqnarray}
Here, by definition, it is assumed that ${z_{i_{M +1}}} \equiv {z_{i_{1}}}$ ($i=1,2$)
and $q_{M +1} \equiv q_{1}$.
Substituting
\begin{equation}
\label{45}
\frac{1}{
{z_{1}}_{\alpha} + {z_{2}}_{\alpha} - i q_{\alpha} +
{z_{1}}_{\alpha +1}  + {z_{2}}_{\alpha +1}  + i q_{\alpha +1}}
\; = \;
\int_{0}^{\infty} d\omega_{\alpha}
\exp\Bigl[-\bigl(
{z_{1}}_{\alpha} + {z_{2}}_{\alpha} - i q_{\alpha} +
{z_{1}}_{\alpha +1}  + {z_{2}}_{\alpha +1}  + i q_{\alpha +1}
\bigr) \, \omega_{\alpha}
\Bigr]
\end{equation}
into eq.(\ref{44}), we obtain
\begin{equation}
\label{46}
\mbox{Tr} \, \hat{A}^{M} \; =  \;
\int_{0}^{\infty} d\omega_{1} \, ... \, d\omega_{M} \,
\prod_{\alpha=1}^{M}
\Bigl[ 
A\bigl(\omega_{\alpha}, \omega_{\alpha +1}\bigr)
\Bigr]
\end{equation}
where
\begin{eqnarray}
 \nonumber
A\bigl(\omega, \omega'\bigr) & = &
\int_{-\infty}^{+\infty} \frac{dy dq}{2\pi}
\frac{d{\xi_{1}} d{\xi_{2}}d{\xi_{3}}}{(2\pi)^{3/2}} \; 
\Ai\bigl(y + q^{2} + \omega + \omega' - i{\xi_{3}} \sqrt{x}\bigr)
\times
\\
\nonumber
\\
\nonumber
&\times&
\exp\Bigl\{
-\frac{1}{2} \bigl({\xi_{1}} + \frac{1}{2}iq\sqrt{x}\bigr)^{2} 
-\frac{1}{2} \bigl({\xi_{2}} - \frac{1}{2}iq\sqrt{x}\bigr)^{2}
-\frac{1}{2} {\xi_{3}}^{2}
- i q (\omega - \omega')
\Bigr\} 
\times 
\\
\nonumber
\\
\nonumber
&\times&
\int\int_{{\cal C}'}
\frac{d{z_{1}}d{z_{2}}}{(2\pi i)^{2}}
\Bigl(
\frac{2\pi i}{{z_{1}}} \delta({z_{2}}) +
\frac{2\pi i}{{z_{2}}} \delta({z_{1}}) -
\frac{1}{{z_{1}}{z_{2}}}
\Bigr)
\times
\\
\nonumber
\\
&\times&
\exp\Bigl\{
{z_{1}}\bigl(y + f_{1} + {\xi_{1}}\sqrt{x}\bigr) +
{z_{2}}\bigl(y + f_{2} + {\xi_{2}}\sqrt{x}\bigr)
\Bigr\}
\label{47}
\end{eqnarray}
Integrating over $z_{1}$ and $z_{2}$ we get:
\begin{eqnarray}
 \nonumber
A\bigl(\omega, \omega'\bigr) & = &
\int_{0}^{+\infty} dy \int_{-\infty}^{+\infty} \frac{dq}{2\pi}
\Ai\bigl(y + q^{2} - f_{1} + \omega + \omega' + \frac{1}{2} i q x \bigr)
\exp\bigl\{ -iq(\omega-\omega')\bigr\} \; +
\\
\nonumber
\\
\nonumber
&+&
\int_{0}^{+\infty} dy \int_{-\infty}^{+\infty} \frac{dq}{2\pi}
\Ai\bigl(y + q^{2} - f_{2} + \omega + \omega' - \frac{1}{2} i q x \bigr)
\exp\bigl\{ -iq(\omega-\omega')\bigr\} \; +
\\
\nonumber
\\
\nonumber
&-&
\int_{-\infty}^{+\infty} dy \int_{-\infty}^{+\infty} \frac{dq}{2\pi}
\int\int\int_{-\infty}^{+\infty}\frac{d{\xi_{1}} d{\xi_{2}}d{\xi_{3}}}{(2\pi)^{3/2}} \; 
\Ai\bigl(y + q^{2} + \omega + \omega' - i{\xi_{3}} \sqrt{x}\bigr)
\times
\\
\nonumber
\\
\nonumber
&\times&
\exp\Bigl\{
-\frac{1}{2} \bigl({\xi_{1}} + \frac{1}{2}iq\sqrt{x}\bigr)^{2} 
-\frac{1}{2} \bigl({\xi_{2}} - \frac{1}{2}iq\sqrt{x}\bigr)^{2}
-\frac{1}{2} {\xi_{3}}^{2}
- i q (\omega - \omega')
\Bigr\}
\times
\\
\nonumber
\\
&\times&
\theta\bigl(y + f_{1} + \xi_{1}\sqrt{x}\bigr) \; 
\theta\bigl(y + f_{2} + \xi_{2}\sqrt{x}\bigr)
\label{48}
\end{eqnarray}
where $\theta(y)$ is the step function.
Redefining, $\xi_{1} = (t-\eta)/\sqrt{2}, \; $ $\xi_{2} = (t+\eta)/\sqrt{2} , \; $
$\xi_{3} = (it +\zeta)/\sqrt{2}, \; $ and integrating over $q$, $t$ and $\zeta$,
we find the following result:
\begin{eqnarray}
\nonumber
A(\omega,\omega') &=& 
2^{1/3} K\Bigl[2^{1/3}\bigl(\omega - \tilde{f}_{1}\bigr) \, , \; 
               2^{1/3}\bigl(\omega' - \tilde{f}_{1} \bigr)\Bigr]
\exp\Bigl\{\frac{1}{4}(\omega - \omega') x \Bigr\} \; 
+
\\
\nonumber
\\
\nonumber
&+&
2^{1/3} K\Bigl[2^{1/3}\bigl(\omega - \tilde{f}_{2}\bigr) \, , \; 
               2^{1/3}\bigl(\omega' - \tilde{f}_{2} \bigr) \Bigr]
\exp\Bigl\{-\frac{1}{4}(\omega - \omega') x \Bigr\} \; 
-
\\
\nonumber
\\
\nonumber
&-& 2^{2/3}
\int_{-\infty}^{+\infty} dy \;
\int_{-\infty}^{+\infty} \frac{d\eta}{\sqrt{2\pi}}
\Ai\Bigl[2^{1/3}\Bigl(y + \omega - \eta \sqrt{\frac{x}{8}} \;\Bigr)\Bigr]
\Ai\Bigl[2^{1/3}\Bigl(y + \omega' + \eta \sqrt{\frac{x}{8}} \;\Bigr)\Bigr]
\times
\\
\nonumber
\\
&\times&
\exp
\Bigl\{
-\frac{1}{2} \eta^{2} \, - \, \frac{1}{2} x y \, - \, \frac{1}{4} x (\omega + \omega')
+\frac{1}{3}  \Bigl(\frac{x}{4}\Bigr)^{3} 
\Bigr\}
\;
\theta\Bigl(y + \tilde{f}_{1} - \eta \sqrt{\frac{x}{8}} \; \Bigr) \; 
\theta\Bigl(y + \tilde{f}_{2} + \eta \sqrt{\frac{x}{8}} \; \Bigr) 
\label{51}
\end{eqnarray}
where $\tilde{f}_{1,2} = \frac{1}{2}\bigl(f_{1,2} - x^{2}/16 \bigr)$ and
 $K\bigl(\omega, \omega'\bigr) = \int_{0}^{\infty} dy \Ai(y + \omega)\Ai(y + \omega')$
is the Airy kernel.

Thus the distribution function $W(f_{1}, f_{2}; x)$, eq.(\ref{5}),
is given the Fredholm determinant
\begin{equation}
 \label{52}
W(f_{1}, f_{2}; x) \; = \; \det\bigl[1 \, - \, \hat{A} \bigr]
\end{equation}
where $\hat{A}$ is the integral operator with the kernel 
$A(\omega,\omega') \; \; \; (\omega,\omega' \geq 0)$
given in eq.(\ref{51}).

Note that using explicit expression (\ref{51}) one can easily test the
obtained result for three limit cases:
\begin{eqnarray}
 \label{53}
\lim_{f_{1}\to -\infty}
A(\omega,\omega') &=& 
2^{1/3} K\Bigl[2^{1/3}\bigl(\omega - \tilde{f}_{2}\bigr) \, , \; 
               2^{1/3}\bigl(\omega' - \tilde{f}_{2} \bigr) \Bigr]
\exp\Bigl\{-\frac{1}{4}(\omega - \omega') x \Bigr\}
\\
\nonumber
\\
\label{54}
\lim_{f_{2}\to -\infty}
A(\omega,\omega') &=&
2^{1/3} K\Bigl[2^{1/3}\bigl(\omega - \tilde{f}_{1}\bigr) \, , \; 
               2^{1/3}\bigl(\omega' - \tilde{f}_{1} \bigr)\Bigr]
\exp\Bigl\{\frac{1}{4}(\omega - \omega') x \Bigr\}
\\
\nonumber
\\
\nonumber
\lim_{x\to 0}
A(\omega,\omega') &=&
2^{1/3} K\Bigl[2^{1/3}\bigl(\omega - f_{1}/2\bigr) \, , \; 
               2^{1/3}\bigl(\omega' -f_{1}/2\bigr)\Bigr] \; \theta (f_{1} - f_{2})  \; +
\\
\nonumber
\\
&+&
2^{1/3} K\Bigl[2^{1/3}\bigl(\omega - f_{2}/2\bigr) \, , \; 
               2^{1/3}\bigl(\omega' -f_{2}/2\bigr)\Bigr] \; \theta (f_{2} - f_{1}) 
\label{55}
\end{eqnarray}
which demonstrate that in the case $f_{1,2} \to -\infty$ we recover the usual GUE Tracy-Widom
distribution for $f_{2,1}$ correspondingly, while in the limit case $x \to 0$ we find
the usual GUE Tracy-Widom distribution for $f_{1}$ (in the case $f_{1} > f_{2}$) and 
for $f_{2}$ (in the case $f_{2} > f_{1}$), as it should be.

\section{Conclusions}
 
 In view of the recent proof \cite{Imamura-Sasamoto-Spohn} that the result of the present calculations 
is equivalent to the  one obtained earlier by Prolhac and Spohn \cite{Prolhac-Spohn} we can conclude that 
the Bethe ansatz replica technique has demonstrated  (once again) its the efficiency and robustness
which allows to perform computations of sufficiently complicated objects in rather simple way.

\acknowledgments

I am grateful to Alexei Borodin and  Herbert Spohn for numerous illuminating discussions.




\begin{thebibliography}{99}









\bibitem{Prolhac-Spohn} S.\ Prolhac and H.\ Spohn,
         J.Stat.Mech. P01031 (2011)

\bibitem{KPZ-TW1a} T.Sasamoto and H.Spohn,
         Phys.\ Rev.\ Lett.\ {\bf 104}, 230602 (2010)

\bibitem{KPZ-TW2}  G.Amir, I.Corwin and J.Quastel,
        Comm.\ Pure Appl.\ Math.\ {\bf 64}, 466 (2011)

\bibitem{BA-TW2} V.Dotsenko,
         EPL, {\bf 90},20003 (2010)

\bibitem{LeDoussal1} P.Calabrese, P. Le Doussal and A.Rosso,
         EPL, {\bf 90},20002 (2010);


\bibitem{Imamura-Sasamoto-Spohn} T.Imamura, T.Sasamoto and H.Spohn,
{\it On the equal time two point distribution of the 1D KPZ equation by replica},
 arXiv:1305.1217 (2013)



\end{thebibliography}
\end{document}